H. Che

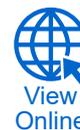

View Online

Export Citation

# Plasma compressibility and the generation of electrostatic electron Kelvin–Helmholtz instability

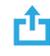

H. Che[a)]

AFFILIATIONS

Center for Space Plasma and Aeronomic Research (CSPAR), University of Alabama in Huntsville, Huntsville, Alabama 35805, USA
and Department of Space Science, University of Alabama in Huntsville, Huntsville, Alabama 35899, USA

[a)]Author to whom correspondence should be addressed: hc0043@uah.edu

## ABSTRACT

This study explores the generation of electrostatic (ES) electron Kelvin–Helmholtz instability (EKHI) in collisionless plasma with a step-function electron velocity shear akin to that developed in the electron diffusion region in magnetic reconnection. In incompressible plasma, ES EKHI does not arise in any velocity shear profile due to the decoupling of the electric potential from the electron momentum equation. Instead, a fluid-like Kelvin–Helmholtz instability (KHI) can arise. However, in compressible plasma, the compressibility couples the electric potential with the electron dynamics, leading to the emergence of a new ES mode EKHI on Debye length $\lambda_{De}$, accompanied by the co-generation of an electron acoustic-like wave. The minimum threshold of ES EKHI is $\Delta U > 2c_{se}$, i.e., the electron velocity shear is larger than twice the electron acoustic speed $c_{se}$. The corresponding growth rate is $Im(\omega) = ((\Delta U/c_{se})^2 - 4)^{1/2}\omega_{pe}$, where $\omega_{pe}$ is the electron plasma frequency.

## I. INTRODUCTION

Kelvin–Helmholtz instability (KHI)[1] is one of the most common instabilities in space plasma observations. KHI is driven by velocity shear in a single continuous fluid or a velocity difference across the interface between two fluids and can also occur in the magnetohydrodynamics (MHD) framework.[2–6] KHI has been treated as a large-scale fluid instability, and its importance on kinetic scales has not attracted sufficient attention until recently, when both observations and particle-in-cell (PIC) simulations have discovered that on electron kinetic-scale, the electron Kelvin–Helmholtz instability (EKHI) driven by the electron velocity shear plays an important role in electron acceleration, indicating that the kinetic-scale KHI may have important applications in planetary magnetospheric substorms and solar flares.[7–14]

Magnetic reconnection is believed to be responsible for the fast release of magnetic energy in space and astrophysical environments. In magnetic reconnection, the current sheet must thin to the electron inertial length $d_e$ to break the electron frozen-in condition $\mathbf{E} + \mathbf{v}_e \times \mathbf{B}/c = 0$. It is common that anti-parallel electron streaming along the anti-parallel magnetic field lines also develops and triggers EKHI, resulting in strong electron acceleration.[12,15–17] Recent *Magnetospheric Multiscale* (MMS) observations appear to have identified EKHI and the corresponding vortical electron acceleration in magnetospheric magnetic reconnection.[10,11,14,18]

On the MHD spatial scale, ions and electrons are treated as a single fluid and the MHD Ohm's law and the momentum equation are independent. On electron dynamic scales $\sim d_e$, ions decouple from electrons, electron dynamics dominate, and ions are usually approximated as the background. In this case, the electron momentum equation is also the generalized Ohm's law.[19] Our recent study of the electromagnetic (EM) EKHI[20] showed that the dispersion relation of the EM EKHI in a step-function velocity shear is similar to that of the ideal MHD KHI if we assume the conservation of magnetic flux. In other words, the EM mode KHIs are similar on different scales.

In addition to the EM mode, which is caused by the couplings between the velocity shear and Faraday's and Ampère's laws, there is an electrostatic (ES) mode of KHI on the electron dynamic scale, which is caused by the coupling between the velocity shear and the Poisson equation. Different from the EM mode, the charge separation in ES mode can only occur on the kinetic scales, and the kinetic effects must be studied using either two-fluid equations or the Vlasov equation.[2,21–24]

In this paper, we investigate the ES EKHI using the electron fluid equations coupled with the Poisson equation for both incompressible and compressible collisionless plasma, treating ion fluid as a background. We investigate how compressibility affects the onset of the ES EKHI. We show that in incompressible plasma, the electron potential decouples from the electron momentum equation, and only a pure electron fluid-like KHI can arise. In contrast, in the compressible plasma, we found that the compressibility couples the electric potential and the charge separation with the electron momentum equation, leading to an ES EKHI with the wavelength on the Debye length and the frequency on electron plasma frequency. Simultaneously, an electron acoustic-like wave is also generated along with the ES EKHI.

## II. ELECTRON DYNAMIC EQUATIONS FOR ES EKHI

Electron dynamic scales run from the Debye length $\lambda_{De} \equiv v_{te}/\omega_{pe}$ to the electron inertial length $d_e \equiv c/\omega_{pe}$, where $v_{te}$ is the electron thermal speed and $\omega_{pe}$ is the electron plasma frequency. On the electron dynamic scale, electrons decouple from ions, and the high-frequency electron dynamics dominate and the low-frequency ion dynamics is neglected in our calculations. Different from the ES EKHI in strong magnetized fusion plasma[25,26] in which the electron velocity is generated due to $\mathbf{E} \times \mathbf{B}$ drift, during the thinning of the current sheet in a magnetic reconnection to the electron inertial length $d_e$, the in-plane electron velocity shear can develop following the anti-parallel reconnection magnetic field in the electron diffusion region, especially in force-free magnetic reconnection, where the strong velocity shear triggers EM EKHI.[12,17] In addition, in the electron diffusion region of magnetic reconnection, the in-plane magnetic field is very weak and is close to zero around the middle plane. The out-of-plane magnetic field caused by the anti-parallel electron current density is approximately proportional to the distance to the center of the current sheet estimated from the Ampère's law $\nabla \times \mathbf{B} = \frac{4\pi}{c}\mathbf{j}_e$, and it approaches zero as drawing near to the center of the current sheet. We neglect the out-of-plane magnetic field component in this study since the EKHI occurs in the neighborhood of the center of the current sheet. The initial pressure $P_0$ is determined by the initial equilibrium $P_0 + B_0^2/8\pi = constant$ on both sides of $z = 0$. Thus, for simplicity, we assume initially the magnetic field and electric field are zeros. With these assumptions, we have the following equations:

$$\partial_t n_e + \nabla \cdot (n_e \mathbf{v}_e) = 0, \quad (1)$$

$$m_e n_e (\partial_t + \mathbf{v}_e \cdot \nabla) \mathbf{v}_e - e n_e \nabla \phi + \nabla P_e = 0, \quad (2)$$

and the system couples with Poisson equations

$$\nabla^2 \phi = 4\pi e (n_e - n_i), \quad (3)$$

where $\phi$ is the electric potential.

We linearize these equations assuming the unperturbed ion and electron density is $n_{i0} = n_{e0} = n_0$, the initial electron pressure is $P_0$, and the initial velocity $\mathbf{U}_0 = U_0(z)\hat{x}$ is along the $x$-direction and is a function of the height $z$ from the interface at $z = 0$, where the velocity shear is discontinuous. Treating the ions as a stationary background and let $\mathbf{v}_e = \mathbf{U}_0 + \delta\mathbf{v}$, $n_e = n_0 + \delta n$, $P_e = P_0 + \delta P$ and $\phi = \delta\phi$, we have

$$\partial_t \delta n + U_0 \partial_x \delta n + n_0 \nabla \cdot \delta\mathbf{v} + \delta v_z \partial_z n_0 = 0, \quad (4)$$

$$m_e n_0 (\partial_t + U_0 \partial_x)\delta\mathbf{v} + m_e n_0 \delta v_z \partial_z \mathbf{U}_0 - e n_0 \nabla \delta\phi + \nabla \delta P = 0, \quad (5)$$

$$\nabla^2 \delta\phi = 4\pi e \delta n. \quad (6)$$

## III. THE SUPPRESSION OF ELECTROSTATIC ELECTRON KELVIN–HELMHOLTZ INSTABILITY IN INCOMPRESSIBLE PLASMA

In the incompressible plasma, the incompressibility condition $\nabla \cdot \mathbf{v} = 0$ provides an additional constraint for the motion of electrons. With this restriction, in the following, we show that the electric potential is not involved in the generation of ES EKHI, and as a result, the EKHI in incompressible plasma behaves like a fluid KHI.

The linearized incompressibility condition is

$$\nabla \cdot \delta\mathbf{v} = 0. \quad (7)$$

Assuming the fluctuations have the form $\delta f(x,y,z) = \delta f(z)e^{i(k_x x + k_y y - \omega t)}$, we rewrite the component equations for the linearized incompressibility condition (7) and the electron momentum equation (5) as

$$k_x \delta v_x + k_y \delta v_y - i\partial_z \delta v_z = 0, \quad (8)$$

$$-i m_e n_0 \Omega \delta v_x + m_e n_0 \delta v_z \partial_z U_0 - i k_x e n_0 \delta\phi + i k_x \delta P = 0, \quad (9)$$

$$-i m_e n_0 \Omega \delta v_y - i k_y e n_0 \delta\phi + i k_y \delta P = 0, \quad (10)$$

$$-i m_e n_0 \Omega \delta v_z - e n_0 \partial_z \delta\phi + \partial_z \delta P = 0. \quad (11)$$

Multiplying Eq. (9) by $ik_x$ and Eq. (10) by $ik_y$, and adding them up, then applying Eq. (8), we have

$$k^2 \delta P = i m_e n_0 \Omega \partial_z \delta v_z + i m_e n_0 \delta v_z \partial_z U_0 + k^2 e n_0 \delta\phi, \quad (12)$$

where $k^2 = k_x^2 + k_y^2$. Inserting Eq. (12) into Eq. (11), we obtain

$$k^2 n_0 \Omega \delta v_z - \partial_z [n_0 \Omega \partial_z \delta v_z + n_0 \delta v_z \partial_z U_0] = 0, \quad (13)$$

where $\Omega = \omega - k_x U_0$.

Equation (13) is similar to the equation obtained for the fluid-like KHI by Chandrasekhar ignoring the gravity and surface tension.[1] With a given velocity shear function $\mathbf{U}_0$ and density $n_0$, Eq. (13) leads to the dispersion relation for electron fluid KHI, which is independent of the electric potential $\phi$.

Since the electron density fluctuation is connected to the electric potential, we want to know whether the fluid-like KHI generated density fluctuations can produce an electric potential growth, and how the electric potential affects the fluid-like KHI.

Let us assume both the $\mathbf{U}_0$ and density $n_0$ have step-function profiles

$$\mathbf{U}_0 = \begin{cases} U_1 \hat{x}, & z > 0, \\ U_2 \hat{x}, & z < 0, \end{cases} \quad (14)$$

and

$$\mathbf{n}_0 = \begin{cases} n_1, & z > 0, \\ n_2, & z < 0. \end{cases} \quad (15)$$

Integration of Eq. (13) over $z = 0$ yields the famous fluid KHI-like dispersion relation

$$n_1 \Omega_1^2 + n_2 \Omega_2^2 = 0 \quad (16)$$

or

$$\omega = \frac{n_1 \mathbf{k} \cdot \mathbf{U}_1 + n_2 \mathbf{k} \cdot \mathbf{U}_2}{n_1 + n_2} \pm \frac{i}{n_1 + n_2}(n_1 n_2 (\mathbf{k} \cdot \Delta \mathbf{U})^2)^{1/2}, \quad (17)$$

where $\Delta \mathbf{U} = \mathbf{U}_1 - \mathbf{U}_2$. The growth rate $Im(\omega)$ reaches its maximum $|\mathbf{k} \cdot \Delta \mathbf{U}|/2$ when $n_1 = n_2$.

At $z \neq 0$, we have the equation for the electric potential $\delta\phi$

$$\nabla^2 \delta\phi = 0, \quad (18)$$

where $\nabla^2 = \partial_z^2 - k^2$. The solution is

$$\delta\phi(z) = \begin{cases} Ae^{-kz}, & z > 0, \\ Be^{kz}, & z < 0. \end{cases} \quad (19)$$

$\delta\phi$ being continuous at $z = 0$ requires $A = B$, and $\partial_z \delta\phi$ satisfies

$$\partial_z \delta\phi|_{z+0} - \partial_z \delta\phi|_{z-0} = \sigma_e = -2Ak, \quad (20)$$

where $\sigma_e$ is the surface charge density.

From Eq. (4), we have $i\Omega \delta n = \delta v_z \partial_z n_0$, and the Poisson equation becomes

$$(\partial_z^2 - k^2)\delta\phi = 4\pi e \frac{\delta v_z}{i\Omega} \partial_z n_0. \quad (21)$$

For the step-function velocity shear in the incompressible plasma,[20] we have $\nabla^2 \delta v_z = 0$, and the solution is $\delta v_z(z) = C_1 e^{-kz}$ for $z > 0$ and $\delta v_z(z) = C_2 e^{kz}$ for $z < 0$, where $C_1$ and $C_2$ are two constants and $\delta v_z$ is not continuous at the interface at $z = 0$. However, the Lagrangian displacement of the interface $\delta \mathbf{l}$ is continuous, and it is related to $\delta \mathbf{v}$ by $\delta \mathbf{v} = d\delta \mathbf{l}/dt = -i\Omega \delta \mathbf{l}$; thus, we rewrite the solution of $\delta v_z$ to

$$\delta v_z = -i\Omega \delta l_z = \begin{cases} -i\Omega_1 C e^{-kz}, & z > 0, \\ -i\Omega_2 C e^{kz}, & z < 0. \end{cases} \quad (22)$$

Inserting $\delta v_z$ into Eq. (21) yields

$$(\partial_z^2 - k^2)\delta\phi = -4\pi e \delta l_z \partial_z n_0. \quad (23)$$

Integrating the above equation over $z = 0$, we obtain

$$\partial_z \delta\phi|_{z+0} - \partial_z \delta\phi|_{z-0} = -2Ak = -4\pi eC(n_1 - n_2). \quad (24)$$

Thus, we have the magnitude $A$ of $\delta\phi$ and the surface electron charge density $\sigma_e$

$$A = \frac{2\pi eC(n_1 - n_2)}{k}, \quad (25)$$

$$\sigma_e = 4\pi eC(n_1 - n_2), \quad (26)$$

and

$$\delta\phi(z) = \begin{cases} \dfrac{2\pi eC(n_1 - n_2)}{k}e^{-kz}, & z > 0, \\ \dfrac{2\pi eC(n_1 - n_2)}{k}e^{kz}, & z < 0. \end{cases} \quad (27)$$

When $n_1 = n_2$, the growth rate of the fluid-like KHI reaches its maximum $Im(\omega) = |\mathbf{k} \cdot \Delta \mathbf{U}|/2$. However, Eqs. (25) and (26) tell us that neither the surface electron charge density $\sigma_e$ nor the electric potential $\delta\phi$ is produced at the interface at $z = 0$ since the KHI driven mass flows $m_e n_1 \delta v_z$ and $m_e n_2 \delta v_z$ bring in or out the same amount of electrons to/from the two sides of the interface $z = 0$ due to the incompressibility of the velocity. When $n_1 \neq n_2$, the KHI growth rate decreases with the increase in $|n_1 - n_2|$. However, as a contrast, the KHI driven mass flows bring in/out different amount of electrons from the two sides of $z = 0$, and a surface electron charge density is produced, which is proportional to the electron density difference $\sigma_e \propto n_1 - n_2$. The corresponding electric potential amplitude is also proportional to the electron density difference $\delta\phi \propto n_1 - n_2$. The electric potential needs to be supported by the pressure.

Let us further look at the pressure balance at the boundary. Equation (11) gives

$$\partial_z(\delta P - en_0 \delta\phi) = -im_e n_0 \Omega \delta v_z. \quad (28)$$

Inserting the solution of $\delta v_z$ and integrating over z at $z \neq 0$ yield

$$\begin{cases} \delta P_1 - en_1 \delta\phi = -Cm_e n_1 \Omega_1^2 \dfrac{1}{k} e^{-kz} + C_1, & z > 0, \\ \delta P_2 - en_2 \delta\phi = Cm_e n_2 \Omega_2^2 \dfrac{1}{k} e^{kz} + C_2, & z < 0, \end{cases} \quad (29)$$

where we have applied the condition that $\delta\phi$ is continuous at $z = 0$ and $C_1$ and $C_2$ are two integral constants.

For $n_1 = n_2$, we have $\delta\phi = 0$ and the pressure supports the KHI growth. Since we have assumed no surface tension exists at $z = 0$, and thus $\delta P_1 = \delta P_2$, together with Eq. (16), we can conclude that $C_1 = C_2$.

For $n_1 \neq n_2$, we have at $z = 0$

$$\delta P_1 - en_1 \delta\phi = \delta P_2 - en_2 \delta\phi, \quad (30)$$

and hence

$$\delta P_1 - \delta P_2 = e(n_1 - n_2)\delta\phi. \quad (31)$$

Equations (27) and (31) show that $\delta P$ is continuous at $z = 0$ for $n_1 = n_2$ and no electric field is generated—this result is consistent with the boundary condition for the pressure at the interface $z = 0$ in zero surface tension fluid. When $n_1 \neq n_2$, the KHI generated electron velocity flows lead to the electron accumulation or depletion at the interface $z = 0$. The work done by the electron flows leads to the build-up of the pressure imbalance across $z = 0$, which is proportional to the electron density difference, i.e., $\delta P_1 - \delta P_2 \propto (n_1 - n_2)$. Since the electric potential is not coupled with the KHI, and its growth cannot lead to an increase in trapped electrons. In other words, the growth of the surface charge density; thus, the build-up of the potential can be quickly stopped by the pressure gradient. However, the balance between them may not be maintained since any perturbation can cause electrons to quickly diffuse from the high potential to the low potential and reduce the surface charge $\sigma_e$ and electric potential $\delta\phi$ to zero. Therefore, in principle, the generation of the electric potential and pressure imbalance at $z = 0$ are caused by the density gradient at the boundary rather than the electron fluid KHI. The decoupling between the electric potential and KHI cannot support the growth of the electron potential as it is offset by the pressure gradient.

## IV. ELECTROSTATIC ELECTRON KELVIN–HELMHOLTZ INSTABILITY IN COMPRESSIBLE PLASMA

In the incompressible plasma, we have shown that the electric potential decouples from the electron velocity shear and the electron velocity shear drives a purely fluid KHI. The growth rate of KHI reaches its maximum if $n_1 = n_2$. Now let us consider the compressibility of electron fluid $\nabla \cdot \mathbf{v} \neq 0$. For simplicity, we assume the initial electron density $n_0$ and temperature $T_{e0}$ are uniform in the whole space, i.e., $n_1 = n_2$ and $P_1 = P_2$. Only the initial velocity shear has the step-function profile as shown in Eq. (14). We will show that the compressibility couples the electric field with the KHI and generates an ES EKHI and electron acoustic waves on Debye length.

Assuming the perturbation is adiabatic, we have $\delta P = \gamma \delta n_e T_e$, where $\gamma$ is the adiabatic gas constant. Let the perturbations have the form $\delta f = \delta f(z) e^{i(\mathbf{k} \cdot \mathbf{x} - \omega t)}$, then Eqs. (4)–(6) lead to

$$i\Omega \delta n = n_0 \nabla \cdot \delta \mathbf{v}, \quad (32)$$

$$-i\Omega m_e n_0 \delta \mathbf{v} + m_e n_0 \delta v_z \partial_z \mathbf{U}_0 - i e n_0 \mathbf{k} \delta \phi - e n_0 \partial_z \delta \phi \hat{z}$$
$$+ i\gamma T_0 \mathbf{k} \delta n + \gamma T_0 \partial_z \delta n \hat{z} = 0, \quad (33)$$

$$(\partial_z^2 - k^2)\delta \phi = 4\pi e \delta n. \quad (34)$$

The components of the electron momentum equation (33) are

$$-i\Omega m_e n_0 \delta v_x + m_e n_0 \delta v_z \partial_z U_0 - i e n_0 k_x \delta \phi + i\gamma T_0 k_x \delta n = 0, \quad (35)$$

$$-i\Omega m_e n_0 \delta v_y - i e n_0 k_y \delta \phi + i\gamma T_0 k_y \delta n = 0, \quad (36)$$

$$-i\Omega m_e n_0 \delta v_z - e n_0 \partial_z \delta \phi + \gamma T_0 \partial_z \delta n = 0. \quad (37)$$

Multiplying Eq. (35) by $k_x$ and Eq. (36) by $k_y$ and summing them up, then the sum subtracts $\partial_z$ of Eq. (37), yielding the differential equation for $\delta n$ at $z \neq 0$

$$\partial_z^2 \delta n - \left(k^2 + \frac{1}{\lambda_{De}^2} - \frac{\Omega^2}{c_{se}^2}\right)\delta n = 0, \quad (38)$$

where $\lambda_{De}^2 = c_{se}^2/\omega_{pe}^2$, $c_{se} = \sqrt{\gamma T_0/m_e}$ is the electron acoustic speed, and $\omega_{pe}^2 = 4\pi n_0 e^2/m_e$.

The solution of Eq. (38), which is discontinuous at $z \neq 0$, is

$$\delta n = \begin{cases} AFe^{-k_z z}, & z > 0, \\ AFe^{k_z z}, & z < 0, \end{cases} \quad k_z^2 = k^2 + \frac{1}{\lambda_{De}^2} - \frac{\Omega^2}{c_{se}^2}, \quad (39)$$

where $A$ is a constant, and $F$ is a function of $k$ and $\Omega$. We can see that $k_z$ is complex if $\omega$ is complex. Given that $Re(k_z) \lesssim 1/\lambda_{De}$, we have $Re(k_z^2) \lesssim 1/\lambda_{De}^2$. Inserting the solution of $\delta n$ into Eq. (34), we get

$$(\partial_z^2 - k^2)\delta\phi = \begin{cases} 4\pi e A F_1 e^{-k_z z}, & z > 0, \\ 4\pi e A F_2 e^{k_z z}, & z < 0. \end{cases} \quad (40)$$

The general solution of $\delta\phi$ at $z \neq 0$ is

$$\delta\phi = Be^{-k_z|z|} + Ce^{-k|z|}, \quad (41)$$

where $B$ and $C$ are two constants. Inserting this general solution into Eq. (40), we obtain

$$B = \frac{4\pi e A F}{k_z^2 - k^2}, \quad (42)$$

and the constant C is arbitrary.

We know that $\delta\phi$ is continuous at $z = 0$. Thus, $B$ should be a constant, independent of $k_z$ or $\Omega$. Letting $F = k_z^2 - k^2$, we have $B = 4\pi e A$. Thus, we can set $C = 0$ and then

$$\delta\phi = 4\pi e A e^{-k_z|z|}. \quad (43)$$

The $z$-component of the electron momentum equation (33) gives

$$\delta l_z = \frac{1}{m_e n_0}(e n_0 \partial_z \delta\phi - \gamma T_0 \partial_z \delta n). \quad (44)$$

Plugging in the solutions of $\delta\phi$ and $\delta n$ into the above equation, and using the condition that $\delta \mathbf{l}$ is continuous at the interface, i.e., $\delta l_{z1} = \delta l_{z2}$ at $z = 0$, we obtain

$$(\widetilde{k}_{z1}^2 - \widetilde{k}^2)\widetilde{k}_{z1} - \widetilde{k}_{z1} = \widetilde{k}_{z2} - (\widetilde{k}_{z2}^2 - \widetilde{k}^2)\widetilde{k}_{z2}, \quad (45)$$

where $\widetilde{k} \equiv k\lambda_{De}$, $\widetilde{k}_z \equiv k_z \lambda_{De}$, and $\widetilde{\Omega} \equiv \Omega/\omega_{pe}$, thus

$$\widetilde{k}_z^2 = \widetilde{k}^2 + 1 - \widetilde{\Omega}^2. \quad (46)$$

The wavelength is required to be longer than Debye length $\lambda_{De}$, i.e., $Re(\widetilde{k}_z) < 1$. Factorizing Eq. (45) yields

$$\widetilde{k}_{z1} + \widetilde{k}_{z2} = 0, \quad (47)$$

$$\widetilde{k}_{z1}^2 - \widetilde{k}_{z1}\widetilde{k}_{z2} + \widetilde{k}_{z2}^2 - \widetilde{k}^2 - 1 = 0. \quad (48)$$

It is easy to show that Eq. (47) does not have a valid solution, i.e., $\widetilde{k}_{z1} + \widetilde{k}_{z2} \neq 0$. Equation (48) can be approximated by

$$(\widetilde{\Omega}_1^2 + \widetilde{\Omega}_2^2)(\widetilde{\Omega}_1^2 + \widetilde{\Omega}_2^2 - \widetilde{k}^2 - 1) \approx 0 \quad (49)$$

if we neglect the term $\widetilde{\Omega}_1^2 \widetilde{\Omega}_2^2$ in $\widetilde{k}_{z1}\widetilde{k}_{z2}$ due to its amplitude being much smaller than the terms related to $\widetilde{k}_{z1}^2 + \widetilde{k}_{z2}^2$.

Equation (49) results in two dispersion relations

$$\widetilde{\Omega}_1^2 + \widetilde{\Omega}_2^2 = 0, \quad (50)$$

$$\widetilde{\Omega}_1^2 + \widetilde{\Omega}_2^2 - \widetilde{k}^2 - 1 = 0. \quad (51)$$

Equation (50) is the same pure fluid mode as we have seen in the incompressible plasma shown in Eq. (16) for $n_1 = n_2$, which is not a mode of ES EKHI. On the other hand, Eq. (51) reveals a new ES mode that only arises in the compressible plasma

$$\widetilde{\omega} = \frac{\widetilde{\mathbf{k}} \cdot \widetilde{\mathbf{U}}_1 + \widetilde{\mathbf{k}} \cdot \widetilde{\mathbf{U}}_2}{2} + \frac{i}{2}((\widetilde{\mathbf{k}} \cdot \Delta\widetilde{\mathbf{U}})^2 - 2(\widetilde{k}^2 + 1))^{1/2}, \quad (52)$$

where we have ignored the damping mode (negative imaginary mode) using the restriction that $Re(\widetilde{k}_z) < 1$.

If we compare Eq. (52) to Eq. (17) for $n_1 = n_2$, we can see that the only difference in Eq. (52) is a small modification $2(\widetilde{k}^2 + 1)$ added to the imaginary part, but this additional term produces a new kinetic ES EKHI mode. The threshold for the ES EKHI imposed by Eq. (52) is

$$|\widetilde{U}_1 - \widetilde{U}_2| > \sqrt{2\left(1 + \frac{1}{\widetilde{k}^2}\right)}. \quad (53)$$

An interesting property of the function $1 + \frac{1}{\widetilde{k}^2}$ in Eq. (53) is that the threshold decreases with the increase in the wavenumber $\widetilde{k}$, which implies that the maximum growth rate $Im(\omega)$ of ES EKHI mode is reached for wavelengths close to the Debye length, the shortest wavelength of plasma waves. In other words, since $\widetilde{k} \lesssim 1$, the wave mode with $\widetilde{k} \sim 1$ requires the lowest velocity shear to trigger

$$|\Delta \widetilde{\mathbf{U}}| > 2, \quad (54)$$

and the corresponding growth rate $Im(\omega)$ is the Maximum for the same velocity shear

$$Im(\omega) = \frac{1}{2}(\Delta \widetilde{\mathbf{U}}^2 - 4)^{1/2} \omega_{pe}, \quad (55)$$

and the corresponding real frequency of EKHI waves is

$$Re(\omega) = \frac{1}{2}\mathbf{k} \cdot (\widetilde{\mathbf{U}}_1 + \widetilde{\mathbf{U}}_2) \omega_{pe}. \quad (56)$$

The wave with $k \sim \lambda_{De}$ is the most favorable mode in the compressible plasma in which the compressibility couples the pressure, implying that the ES EKHI mode is caused by the coupling between the electron sound wave and the Langmuir wave. In the incompressible plasma, it is unable to generate acoustic wave and thus there is no ES EKHI.

Let us now look at the ES EKHI generated electric potential and the associated ES waves along the z-direction in the compressible plasma. Equation (43) shows that $\delta\phi \propto e^{-k_z|z|}$. The real part of $k_z$ is related to the decay of the ES EKHI waves along $z$, while the imaginary part indicates that the ES mode wave is co-generated propagating along $z$ with a speed near the electron acoustic wave speed $c_{se}$, and this wave grows with the same growth rate $Im(\omega)$ and frequency $Re(\omega)$ shown in Eq. (52). Inserting $\widetilde{\omega}$ for the growth mode in Eq. (52) into $\widetilde{k}_z^2$ in Eq. (46), we have

$$\widetilde{k}_{z1}^2 = \frac{\widetilde{k}^2 + 1}{2} - \frac{i}{2}\widetilde{\mathbf{k}} \cdot (\widetilde{\mathbf{U}}_1 - \widetilde{\mathbf{U}}_2)\left[(\widetilde{\mathbf{k}} \cdot \Delta \widetilde{\mathbf{U}})^2 - 2(\widetilde{\mathbf{k}}^2 + 1)\right]^{1/2}, \quad (57)$$

$$\widetilde{k}_{z2}^2 = \frac{\widetilde{k}^2 + 1}{2} + \frac{i}{2}\widetilde{\mathbf{k}} \cdot (\widetilde{\mathbf{U}}_1 - \widetilde{\mathbf{U}}_2)\left[(\widetilde{\mathbf{k}} \cdot \Delta \widetilde{\mathbf{U}})^2 - 2(\widetilde{\mathbf{k}}^2 + 1)\right]^{1/2}. \quad (58)$$

These equations show that both $k_{z1}^2$ and $k_{z2}^2$ have the same real parts, which are smaller than 1, i.e., $(\widetilde{k}^2 + 1)/2 < 1$. The imaginary parts have the same amplitude but with opposite signs, implying that the waves generated at both sides of the interface at $z = 0$ propagate in the same direction. Assuming $U_1 > U_2$, and let $\gamma = Im((\widetilde{k}_{z2}^2)^{1/2}) = -Im((\widetilde{k}_{z1}^2)^{1/2})$ and $K_z = Re((\widetilde{k}_{z2}^2)^{1/2}) = Re((\widetilde{k}_{z1}^2)^{1/2})$, we have

$$\delta\phi \propto \begin{cases} 4\pi e\, e^{-K_z z} e^{i\gamma z}, & z > 0, \\ 4\pi e\, e^{K_z z} e^{i\gamma z}, & z < 0, \end{cases} \quad (59)$$

and

$$\delta n \propto \begin{cases} (k_{z1}^2 - k^2) e^{-K_z z} e^{i\gamma z}, & z > 0, \\ (k_{z2}^2 - k^2) e^{K_z z} e^{i\gamma z}, & z < 0, \end{cases} \quad (60)$$

where the complex coefficients $(k_z^2 - k^2)$ add a phase to the wave propagation and result in the discontinuity of $\delta n$ at the interface and the generation of $\delta\phi$ due to the velocity shear $\Delta \mathbf{U}$. Similar to $\delta\phi$, $\delta n$ also propagates along $z$. The wavelength is $\sim \lambda_{De}$, and the velocity shear is $\sim c_{se}$; thus, the corresponding phase speed is close to the electron acoustic wave speed $c_{se}$.

Different from the electron scale fluid KHI in the incompressible plasma, the compressibility couples $\delta\phi$ and $\delta n$ with the EKHI to trigger an ES mode; in other words, it is the coupling between the electron acoustic wave and the Langmuir wave. Both the electric potential and the electron charge trapping grow simultaneously with the EKHI in the whole space, while in the incompressible plasma, electron charge density is nonzero only at the interface. The electric potential $\delta\phi$ is supported by the compressibility rather than the pressure which is continuous at $z = 0$ for $n_1 = n_2$.

## V. CONCLUSIONS AND DISCUSSIONS

In this paper, we investigated the onset of ES EKHI in collisionless and inviscid unmagnetized plasma. The configuration we are interested in represents what is commonly seen in the electron diffusion region of space and solar magnetic reconnections, where the antiparallel magnetic field is negligible but a high anti-parallel electron velocity shear is present, e.g., magnetic reconnection in force-free current sheet.[12,17] In the unmagnetized plasma, we do not need to consider the $\mathbf{E} \times \mathbf{B}$ drift whose shear is common in fusion plasma as discussed by Sydora et al.[26]

Unlike the EM EKHI,[20] which has an ideal MHD-like KHI dispersion relation on the electron dynamic scale, ES EKHI does not have an MHD-scale counterpart because the ES instability couples with the Poisson equation and the charge separation becomes important. Such an effect requires a kinetic treatment, or at least using the two-fluid equations. On ion dynamic scales, a macroscopic ES KHI was previously found.[2,21–24] On the electron dynamic scale, ion dynamics can be neglected, and only electron dynamics govern the ES EKHI. Our study shows that, different from the EM EKHI, which can occur in the incompressible plasma, ES EKHI cannot be triggered in the incompressible plasma and compressibility plays an important role in driving the ES EKHI.

In the incompressible and unmagnetized plasma with any electron velocity shear profile, the electron potential decouples from the electron momentum equation and the electron velocity shear drives a pure fluid-like KHI on the electron scale. The KHI generated incompressible velocity flow $\delta v_z$ can bring in/out the electrons to/from the two sides of the interface $z = 0$. For $n_1 = n_2$, no electric field is generated on both sides of the interface due to the same amount of electrons carried by electron flow $\delta v_z$, which is governed by the incompressible condition $\nabla \cdot \delta \mathbf{v} = 0$. For $n_1 \neq n_2$, the electron density difference leads to the electron charge accumulation/depletion at $z = 0$, which is proportional to the electron density difference $n_1 - n_2$. Since the decoupling of electric potential from the KHI, the growth of the potential can not trap more electrons to enhance the further growth of the electric potential since the energy source stored in the velocity shear cannot be converted into the electric field energy directly. As a result, the growth of the electric potential is stopped by the build-up of the pressure gradient at $z = 0$. However, the balance between the pressure and potential is not stable, any perturbation caused by the KHI can break the balance and the electrons can quickly diffuse from the high potential to the low potential, consequently reducing the charge separation and electric potential to zero.

In the compressible plasma, the compressibility couples the electrons' velocity shear and the Poisson equation, which drives a new ES EKHI on Debye length $\lambda_{De}$ and an electron acoustic wave propagating crossing the interface perpendicularly. The minimum threshold for electron velocity shear to trigger an ES EKHI with $k\lambda_{De} \sim 1$ is

$$|\Delta \widetilde{\mathbf{U}}| > 2,$$

while the corresponding growth rate is the maximum for the same velocity shear:

$$Im(\omega) = \frac{1}{2}(\Delta\widetilde{\mathbf{U}}^2 - 4)^{1/2}\omega_{pe}. \qquad (61)$$

where $\widetilde{\mathbf{U}} = \mathbf{U}/c_{se}$, and the corresponding real frequency of EKHI waves is

$$Re(\omega) = \frac{1}{2}\mathbf{k}\cdot(\widetilde{\mathbf{U}}_1 + \widetilde{\mathbf{U}}_2)\omega_{pe}. \qquad (62)$$

The favorable mode with wavelength $\lambda_{De}$ indicates that the ES EKHI is driven by the coupling between the electron acoustic waves and the Langmuir waves which is a topic deserving more investigation. In the incompressible plasma, acoustic wave is not allowed and this can explain the absence of ES EKHI.

Compared to the EM EKHI, in non-magnetized plasma, the velocity shear threshold for triggering the EM EKHI is $\Delta\mathbf{U} > 0$, which is much lower than that for the ES EKHI, implying that the EM EKHI is more likely to be triggered in space plasma, consistent with existing magnetospheric observations. However, it is possible that ES EKHI can be triggered in space plasma if the electron velocity shear is larger than the electron acoustic wave speed. Once ES EKHI is triggered, it can suppress EM EKHI since its growth rate is $\sim \omega_{pe}$, which is $\sim c/v_A$ ($v_A$ is the electron Alfvén wave speed) times larger than that of EM EKHI whose growth rate is $\sim \Omega_{ce}$.

Moreover, we want to emphasize the difference between the ES EKHI driven by a step-profile velocity shear accumulated in magnetic reconnection and the ES EKHI driven by the $\mathbf{E}\times\mathbf{B}$ electron drift in fusion plasma, where an ES EKHI can be generated in the incompressible plasma with a different growth rate.[26] The reason that causes the difference is that the frozen-in condition $\mathbf{E} + \mathbf{v}\times\mathbf{B}/c = 0$ plays an essential role in coupling the electron density and electric potential. In this paper, a simple velocity shear configuration is used. The role of compressibility in more complex situations needs further investigation, in particular, the application in magnetic reconnection requires more sophisticated studies, such as the impact of the guide magnetic field and the reconnection electric field.

### ACKNOWLEDGMENTS

H.C. acknowledged partial support by NSF CAREER 2144324 and a NASA ECIP (Grant No. 80NSSC19K1106).

### AUTHOR DECLARATIONS
#### Conflict of Interest

The authors have no conflicts to disclose.

#### Author Contributions

**H. Che:** Conceptualization (lead); Formal analysis (lead); Writing – original draft (lead); Writing – review & editing (lead).

### DATA AVAILABILITY

Data sharing is not applicable to this article as no new data were created or analyzed in this study.